\documentclass[journal=apchd5,manuscript=article]{achemso}

\usepackage[version=3]{mhchem} % Formula subscripts using \ce{}

\usepackage{graphicx}
\usepackage{placeins}
\graphicspath{{figures/}}
\usepackage{amsmath} % for \dfrac macro
\usepackage{float}
\usepackage{bigints}
\usepackage{color,soul}
\usepackage{booktabs}

\usepackage{xr}
\makeatletter
\newcommand*{\addFileDependency}[1]{
  \typeout{(#1)}
  \@addtofilelist{#1}
  \IfFileExists{#1}{}{\typeout{No file #1.}}
}
\makeatother

\newcommand*{\myexternaldocument}[1]{
    \externaldocument{#1}
    \addFileDependency{#1.tex}
    \addFileDependency{#1.aux}
}
\myexternaldocument{supplemental}

\usepackage{longtable}

\usepackage[T1]{fontenc}
\usepackage{textcomp}
\usepackage{placeins}

\author{Tomi K. Baikie}
\author{James Xiao}
\author{Bluebell Drummond}
\author{Neil C. Greenham}
\email{ncg11@cam.ac.uk}
\author{\\ Akshay Rao}
\email{ar525@cam.ac.uk}
\affiliation[Uni of Cam]
{Cavendish Laboratory, J.J. Thomson Avenue, University of Cambridge, Cambridge, CB3 OHE, UK}

\title[An \textsf{achemso} demo]
  {Optical Efficiency Measurements of Large Area Luminescent Solar Concentrators}

\begin{document}

\begin{tocentry}

  \includegraphics[width=\linewidth]{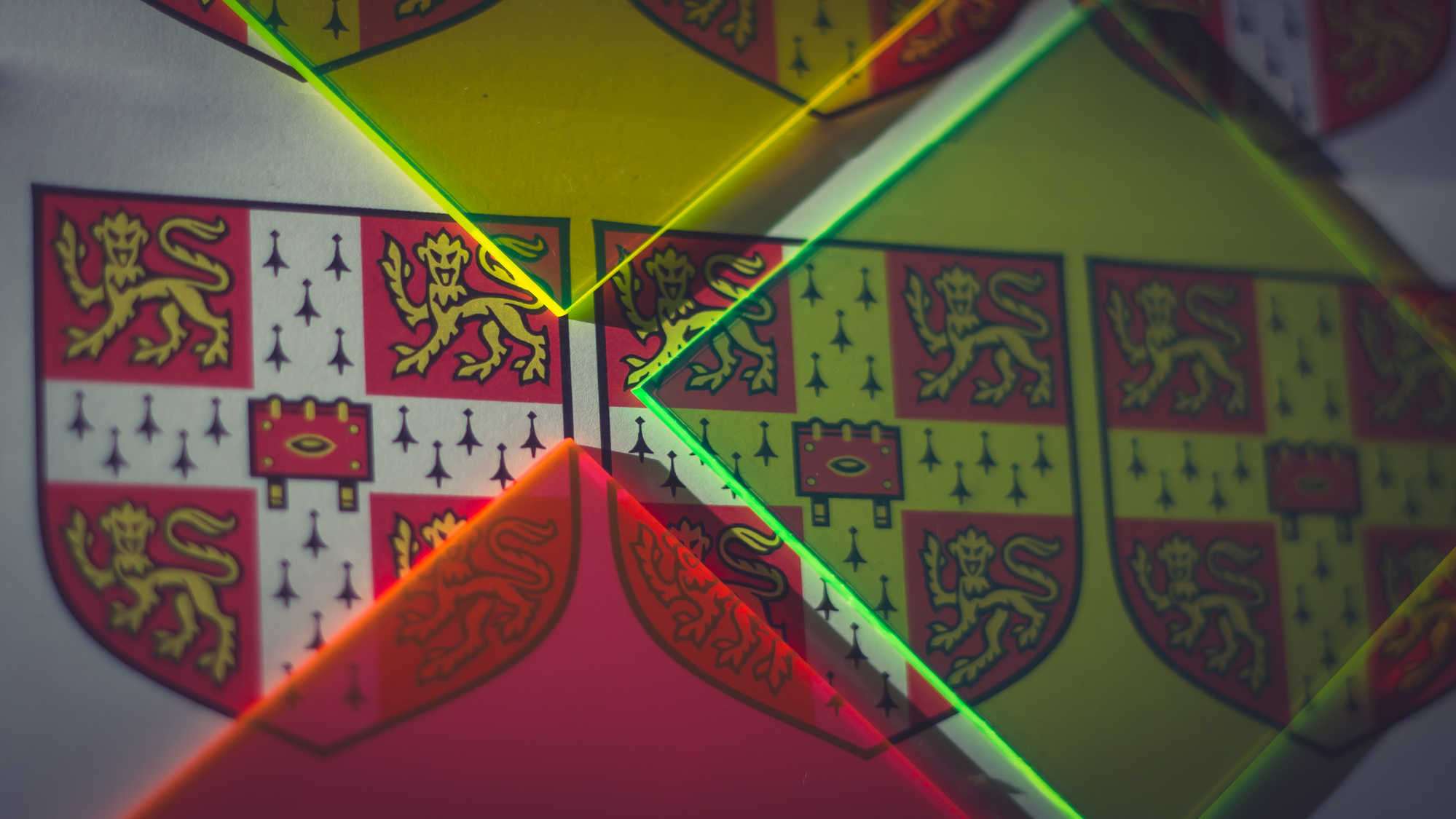}

\end{tocentry}

\begin{abstract}
Luminescent solar concentrators (LSCs) are able to concentrate both direct and diffuse solar radiation and this ability has led to great interest in using them to improve solar energy capture when coupled to traditional photovoltaics (PV). In principle, a large area LSC could concentrate light onto a much smaller area of PV, thus reducing costs or enabling new architectures. However, LSCs suffer from various optical losses which are hard to quantify using simple measurements of power conversion efficiency. Here, we show that spatially resolved photoluminescence quantum efficiency measurements on large area LSCs can be used to  resolve various losses processes such as out-coupling, self-absorption via emitters and self-absorption from the LSC matrix. Further, these measurements allow for the extrapolation of device performance to arbitrarily large LSCs. Our results provide insight into the optimization of optical properties and guide the design of future LSCs for improved solar energy capture.

\end{abstract}

\section{Introduction}
Measurements of the efficiency of luminescent solar concentrators (LSCs) are problematic owing to difficulties in determining re-absorbance, reflectivity, PV characteristics and coupling efficiency, and practical considerations arising from the physical size of the LSC. In the context of LSCs coupled to solar cells, there is great interest in quantitatively establishing the optical and system efficiencies as this provides a means to determine potential improvements in LSC materials and design \cite{Meinardi2017,Shockley1961,Baikie2022ThermodynamicConcentrators}.

\begin{figure}
\centering
  \includegraphics[width=0.5\linewidth]{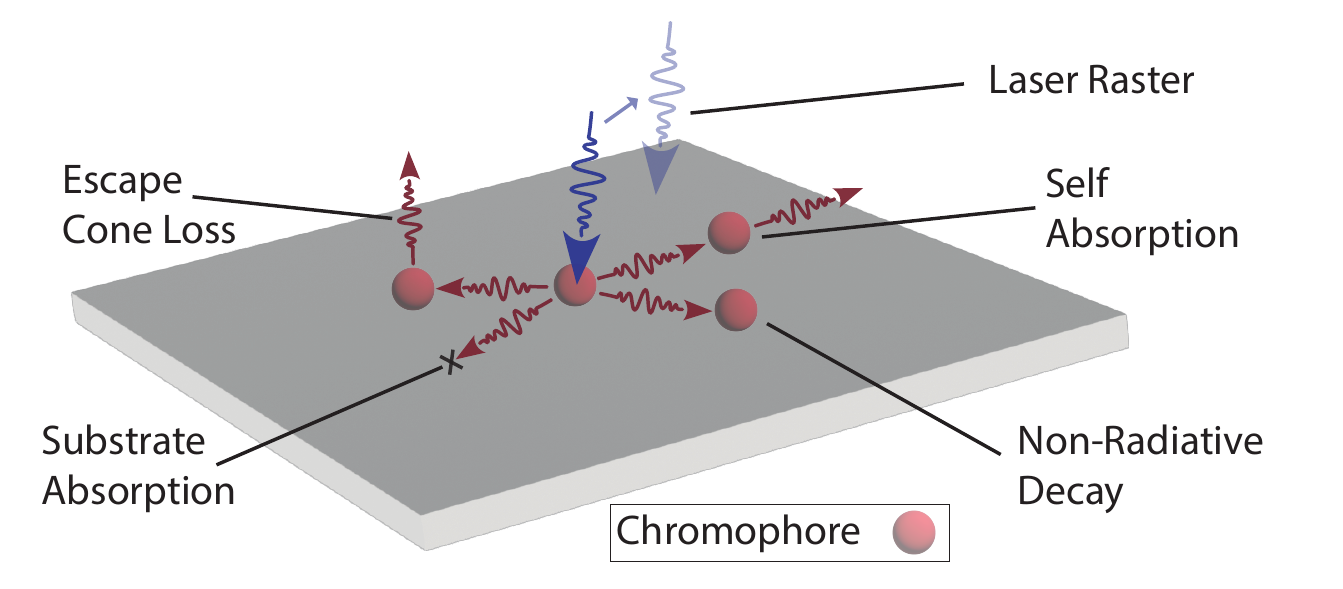}
  \caption{ The laser illumination (in blue) is rastered across the LSC (in grey). Loss mechanisms within the LSC, such as those depicted, have characteristic length scales over which they operate. Measurement of the internal quantum efficiency at different positions of illumination will allow these loss mechanisms to be quantified as a function of LSC size.}
  \label{fig:overview}
\end{figure}

Recent attempts to standardize reporting for LSC device performance are vital, to allow for the direct comparison between different LSC technologies \cite{Yang2022ConsensusPerformance} and highlight the importance of clearly delineating the metrics used to describe the performance of the LSC, typically optical efficiency and power conversion efficiency \cite{Debije2021LaboratoryConcentrators}. However, the reported power conversion efficiency,  $\eta_{\text{dev}}$, may reveal little about the performance of the waveguide itself, since it convolves other factors such as the optical coupling to the solar cell and the properties of the solar cell itself. Therefore, measurements of complete device performance without further quantitative measurements on the optical properties of the LSC itself, will not directly aid our understanding as to which materials and designs are effective, to what extent, and why. It is therefore instructive to understand the optical performance of the LSC itself, before coupling to PV, as this provides a metric to compare LSCs and an understanding of the loss mechanisms in the LSC \cite{Yang2019HowConcentrators,Tummeltshammer2016LossesUnveiled,Wei2019UltrafastConcentrators,You2019EcoFriendlyConcentrators,Wu2018TandemDots}. In this context, measuring the spatially dependent internal quantum efficiency could unravel loss mechanisms of the LSC, as these mechanisms are a function of photon pathlength within the LSC (see \textbf{Figure \ref{fig:overview}}).

The internal quantum efficiency, $\eta_{\text{int}}$, of an LSC is defined by \textbf{Equation \ref{eq:int}},

\begin{equation}\label{eq:int}
    \eta_{\text{int}} = \frac{\textrm{number of photons emitted from edges}}{\textrm{number of photons absorbed by the LSC}}.
\end{equation}

\noindent \textbf{Equation \ref{eq:int}} may be reported for a narrow or broad wavelength range of illumination. Writing \textbf{Equation \ref{eq:int}} in terms of the photon count is most relevant for LSC efficiency, as this directly relates to the number of photogenerated carriers. By measuring the photoluminescence as a function of excitation position, we may determine the $\eta_{\text{int}}$ for arbitrarily large LSCs, and outline how improvements offered by specific technologies will impact LSC efficiencies.

\subsection{Measuring $\eta_{\text{int}}$}
To determine the $\eta_{\text{int}}$, a standard technique involves the use of an integrating sphere \cite{DeMello2004, Greenham1995MeasurementPolymers,deClercq2021ReducingMatrix}. The integrating sphere ($r=25$ cm, Lisun Instruments, in more detail see \textbf{Methods \ref{methods:integratingsphere}}) contains a  coating of a diffusely reflecting material, in our case barium sulphate (Pro-Lite Technology), to ensure that light is redistributed isotropically over the sphere interior regardless of the angle of emission \cite{Walsh1953Photometry}.

 The integrating sphere was calibrated using a NIST-traceable quartz tungsten halogen lamp (Newport 63976 200QC OA) to ensure that both the spectral dependence of reflectance of the sphere is considered as well as any opaque material applied to the sides of the LSC inside the sphere. This results in multiple wavelength-dependent calibration files, detailed in \textbf{SI Section 1.4}.

 To determine $\eta_{\text{int}}$, we follow a revised de Mello method, where a collimated laser beam is directed into the sphere, impinging the LSC \cite{DeMello2004}. As in \textbf{Figure \ref{fig:1}A}, for the first measurement (measurement A) the sphere is empty and laser light alone is measured. The spectral integral of the laser in measurement A is termed $I_{a}$. For the second measurement (measurement B) the LSC is placed inside the sphere and moved out of the beam path so the laser, $I_{b}$,  impinges on the sphere wall. Here, only $\mu$, the fraction of incident laser light scattered by the sphere wall \textit{and} absorbed by the sample will contribute to the laser spectral integral $E_{b}$. For measurement C, the laser is now directed onto the sample and care is taken to ensure the sample is oriented such that reflected laser light from the surface of the sample is directed into the sphere. The spectral integral of the photoluminescence and laser is given by $E_{c}$ and $I_{c}$, respectively. For the fourth measurement, measurement D, opaque material is applied to the edges of the LSC while the laser and sample orientation are the same as in measurement C. The opaque material will prevent emission from the edges of the LSC, leaving only photoluminescence from the front and back faces, given by $E_{d}$. Therefore the efficiency $\eta_{d}$ from measurement D will be the contribution of the entire LSC, $\eta_{c}$, minus the edges, $\eta_{\text{int}}$, contribution, i.e.  $\eta_{d}=\eta_{c}-\eta_{\text{int}}$.

 \textbf{Figure \ref{fig:1}B} depicts the measured spectra from a series of four measurements, with the sharp peak at $405$ nm corresponding to the laser excitation, with the broad profiles at $650$ nm corresponding to the emission of the LSC. As detailed in \textbf{SI Section 1.1}, as long as the LSC is strongly absorbing at the laser wavelength, laser fluctuations are small and calibration corrects for laser absorbance by the opaque material, the expression for internal efficiency simplifies to

\begin{equation}\label{eq:eta}
    \eta_{\text{int}} = \frac{E_{c}-E_{d}}{I_{a}A},
\end{equation}

\noindent where $A$ is the fractional absorption given by $A = (1-{I_{c}}/{I_{b}})$.

\subsection{LSC Size and Self Absorption}

The size of the LSC relative to the integrating sphere introduces a further set of requirements on experimental design. The smallest possible radius of the integrating sphere will produce the highest radiance within the sphere and improve the signal-to-noise ratio of luminescence detection. However, an unavoidable consequence of integrating spheres is the reabsorption of emitted light, which will introduce error in the resulting measured $\eta_{\text{int}}$, which is dependent on the relative geometry between LSC and sphere, as well as the concentration of the chromophore.

To determine the error in $\eta_{\text{int}}$ due to secondary photon reabsorption, we determined the fraction of photons which may contribute to an erroneous signal after being emitted by the LSC. To study how relative sizes of the LSC and integrating sphere relate to $\eta_{\text{int}}$ error, we determined the probability of a photon interacting with the LSC before a photon is measured. The number of times a photon will on average bounce before detection, known as the sphere multiplier, was determined analytically (see \textbf{SI Section 1.5.1} for details).  We determined that a broad range of sphere multipliers approximated a steady state solution for the sphere (see \textbf{SI Section 1.5.1} for details).

We then utilised a Monte Carlo ray tracing algorithm to determine how many photons will interact with the LSC as a function of sphere radius and LSC size. The simulation was run over the number of bounces determined by the sphere multiplier for given LSC and sphere dimensions  (see \textbf{SI Section 1.5.1} for details). We then measured the emission and absorption spectra of the LSC face outside the sphere, identifying the region of overlap between absorption and photoluminescence.  Finally, we analytically determined the probability of reflection or transmission and the associated pathlength of photons impinging isotropically on the LSC, which allowed us to determine what portion of emitted photons may be reabsorbed.

\begin{figure}
\centering
  \includegraphics[width=0.95\linewidth]{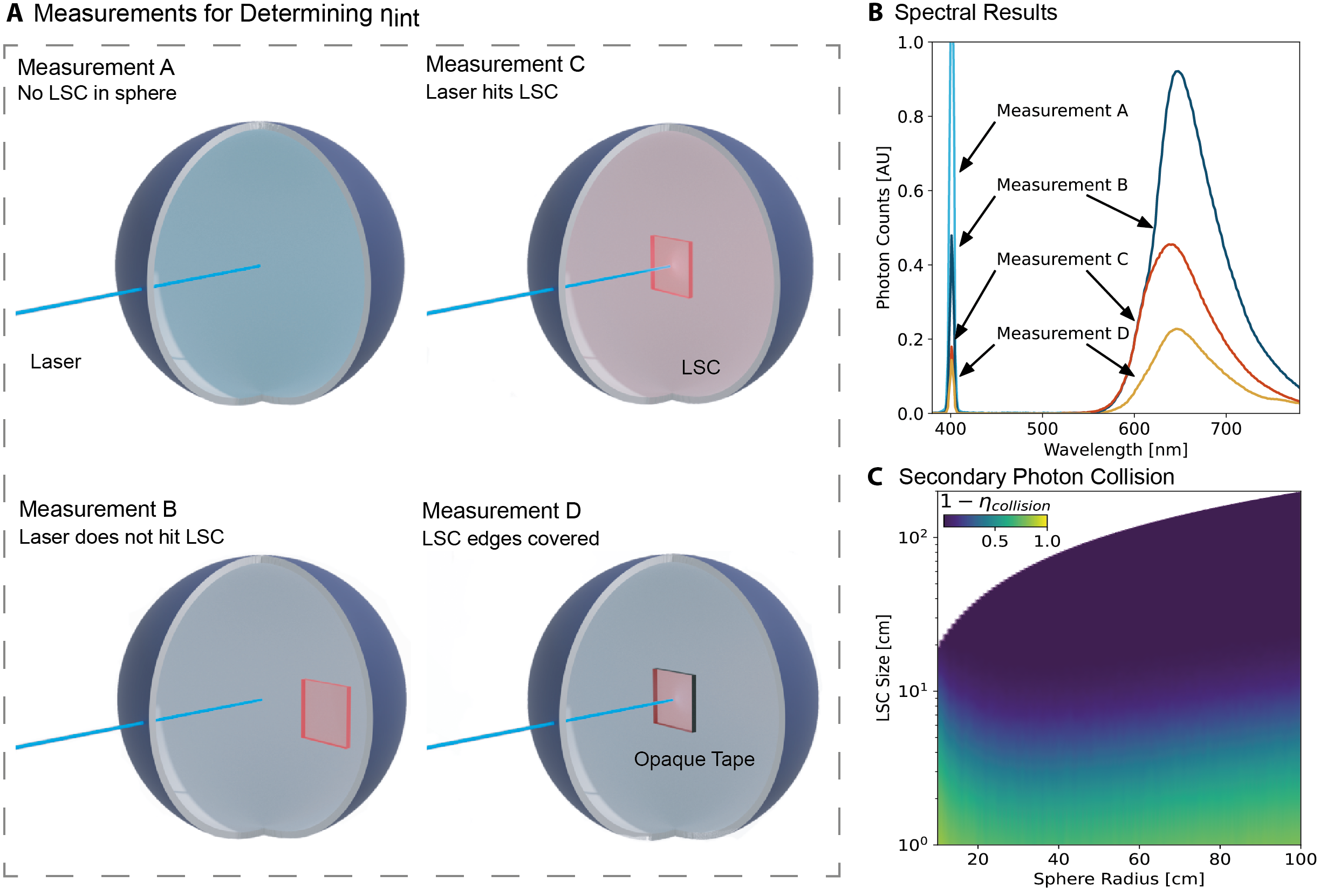}
  \caption{ \textbf{A} -  In all measurements, a laser beam is directed into the integrating sphere through a fibre optic coupling. For measurement A the sphere is empty and laser light alone is measured.  Measurement B the LSC is placed inside the sphere and moved out of the beam path so the laser impinges on the sphere wall. For measurement C the laser is now directed onto the sample.  For measurement D opaque material is applied to the edges of the LSC. \textbf{B} - The spectral results of the 4 measurements to determine $\eta_{\text{int}}$. The emission spectra have been magnified for clarity. \textbf{C} - Photons emitted from LSC intersecting with the LSC before being detected as a function of sphere radius and LSC size.}
  \label{fig:1}
\end{figure}

From these probabilities, we can quantify the relative error in the $\eta_{\text{int}}$ measurement as a function of sphere radius and LSC size for a specific chromophore and concentration. \textbf{Figure \ref{fig:1}C} plots the probability of a photon emitted by the LSC colliding with the LSC for different sphere radii and LSC sizes. Surprisingly, a larger integrating sphere relative to the LSC dimensions does not give a meaningful improvement to the experimental error arising from reabsorption. This is because the average number of photon bounces before detection increases with LSC size, and thus the probability of interaction with the LSC also increases. Typically, the secondary reabsorption error in an $\eta_{\text{int}}$ measurement is dominated by the spectral overlap for all but the smallest LSCs. Minimizing the spectral overlap is a fundamental design goal for LSCs and so, rather usefully, the accuracy of this measurement will increase as LSC chromophores improve \cite{Yablonovitch1980ThermodynamicsConcentrator}.

A detailed analysis of uncertainties and error propagation is given in \textbf{SI Section 1.6}. However, we draw attention here to a few considerations that can have a large impact on the reliability of the measured $\eta_{\text{int}}$. LSCs with an exceptionally low optical density at the excitation wavelength may have an unacceptable level of accuracy using the presented method and may wish to consider the method presented by Yang \textit{et al.}, and determine the $\eta_{\text{dev}}$ alone \cite{Yang2019HowConcentrators}. Additionally, laser fluctuations between the measurements A to D can have a dramatic effect on the calculated $\eta_{\text{int}}$, and as such care should be taken to ensure laser stability. In our case, a power meter (Thorlabs PM16-130) was mounted in the LSC to ensure laser stability before measuring. Laser fluctuations of even 1\% to 5\% over the 4 measurements can induce 50\% fluctuations in the recorded $\eta_{\text{int}}$ for low-absorbance samples.

\subsection{Spatially Resolved Photoluminescence}

\begin{figure}
\centering
  \includegraphics[width=0.95\linewidth]{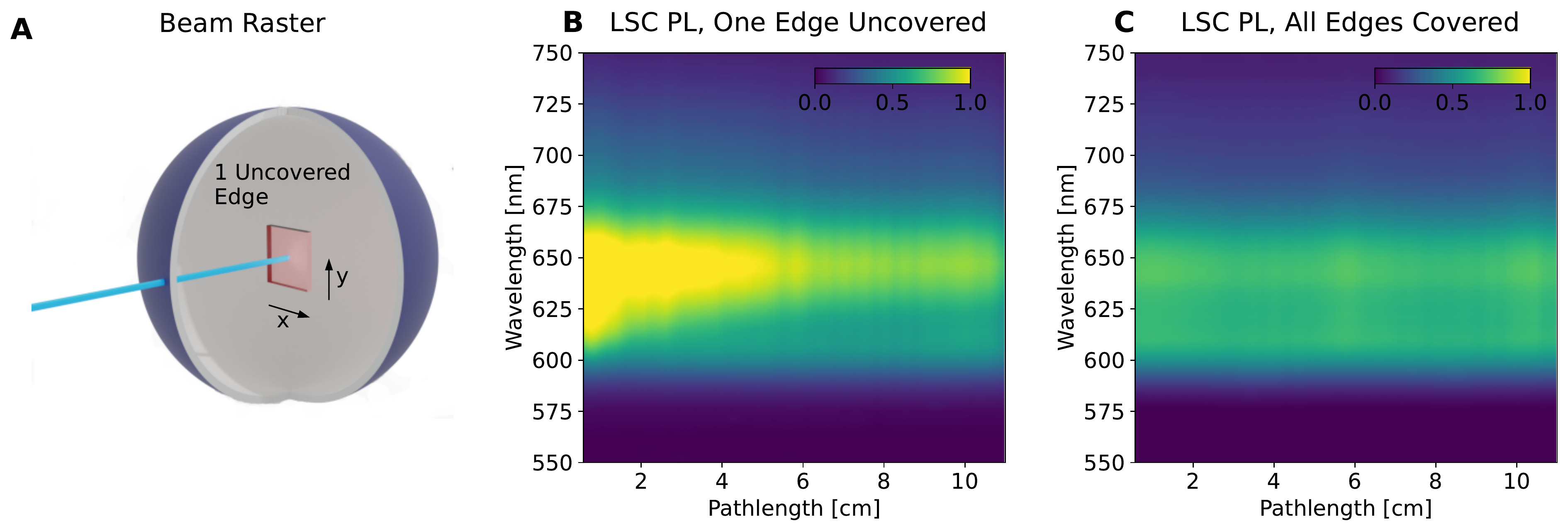}
  \caption{\textbf{A} - The laser beam is rastered across the LSC in the integrating sphere. The pathlength is determined by the average distance for photons to travel from the laser spot to the uncovered edge. \textbf{B} - Photoluminescence as a function of effective pathlength from one edge of the perylene LSC as the laser beam is rastered across the LSC. \textbf{C} -  Photoluminescence as a function of effective pathlength from the front surface alone when all edges of the LSC are covered.
   }
  \label{fig:2}
\end{figure}

Of particular interest in LSC design is $\eta_{\text{int}}$ as a function of the size, or geometric gain, of the LSC. Geometric gain is defined as the ratio between the area of the absorbing face area to the total side area of the LSC perpendicular to illumination. The optical efficiency, $\eta_{\text{int}}$, is unlikely to remain constant with an increasing geometric gain due to additional losses associated with photoluminescence reabsorption or scattering within the LSC \cite{Currie2008High-efficiencyPhotovoltaics}.

By rastering the illumination point over the LSC, as depicted in \textbf{Figure \ref{fig:2}A}, we can determine $\eta_{\text{int}}$ at each point on the LSC, and hence $\eta_{\text{int}}$ as a function of geometric gain within the same experimental setup for arbitrary large LSCs. We recorded photoluminescence as a function of effective pathlength for the square $10 \text{ by} 10 \text{ cm}$ by $3 \text{ mm}$  perylene red LSC (see \textbf{Methods \ref{methods:lsc_manufacture}} for details). Two spatially resolved measurements are required, the first where 3 edges of the LSC are covered (\textbf{Figure \ref{fig:2}B}) and another where all edges are covered (\textbf{Figure \ref{fig:2}C}). The incident laser power ($I_{a}$) and the absorption, as in \textbf{Equation 3}, are the same, so $\eta_{\text{int}}$ can be determined using \textbf{Equation \ref{eq:eta}}. By rastering the point of illumination across the LSC away from the edge of emission, we increase the effective pathlength a photon must travel before it is emitted from the LSC edge.

We distinguish effective pathlength from the actual pathlength travelled by the photon. The pathlength is typically defined as the real-space distance the photon travels within the LSC. This is best determined from LSC ray-tracing simulations \cite{Cambie2017EveryLSC-PMs}. However, we define the effective pathlength here as the average distance from the point of illumination to the emission edge. Although the effective pathlength does not reflect the actual distance the photon travels, it is meaningful in LSC design as it provides a measurable distance over which photon loss occurs. Assuming an isotropic emitter, the effective pathlength, $\overline{l}$ is therefore defined by the length of the paths, $d$, along the angle of acceptance, divided by the angle of acceptance,

\begin{equation}\label{eq:pathlength}
    \overline{l}  =\frac{\int_{\theta_{\min }}^{\theta \max } d(\theta) \mathrm{d} \theta}{\int_{\theta_{\min }}^{\theta \max } 1 \mathrm{~d} \theta}.
\end{equation}

\noindent The analytical solution to \textbf{Equation \ref{eq:pathlength}} is trivial for rectangular LSCs, although the solution is rather lengthy and is therefore detailed in \textbf{SI Section 1.7}.  To determine $\eta_{\text{int}}$ for LSCs of arbitrary size, the photoluminescence must be corrected by a geometric factor to account for the solid angle subtended from the point of illumination to the uncovered edge for the size of LSC. Full derivations of the solid angle correction for arbitrary forms of LSCs are given \textbf{SI Section 1.8}.

In \textbf{Figure \ref{fig:2}B} a spectral shift is readily observed as high-energy photons become redder photons due to chromophore reabsorption and emission within the LSC. The  decay in photoluminescence intensity as a function of pathlength arises from host matrix reabsorption, non-unity PLQE of the chromophore and emission into non-waveguiding modes, known as escape cone losses and also from the change in solid angle of the emission edge as the illumination point is moved across the LSC. \textbf{Figure \ref{fig:2}C}, where all edges are covered and only emission from the top surface is recorded, is then a measure of the photoluminescence from the escape cone. It is not sufficient to use measurement D in \textbf{Figure \ref{fig:1}A}, as the probability of re-absorption and hence escape cone loss may also be a function of distance from the edge. By subtracting the measurement with one edge uncovered by the measurement with all edges obscured, we are left with the photoluminescence coming from the unobscured edge as a function of pathlength from the emitting edge. By extrapolation, we can now determine LSC edge photoluminescence as a function of LSC size, even beyond the size of the measured LSC.

\subsection{Data Analysis} \label{sec:analysis}

\begin{figure}
\centering
  \includegraphics[width=0.95\linewidth]{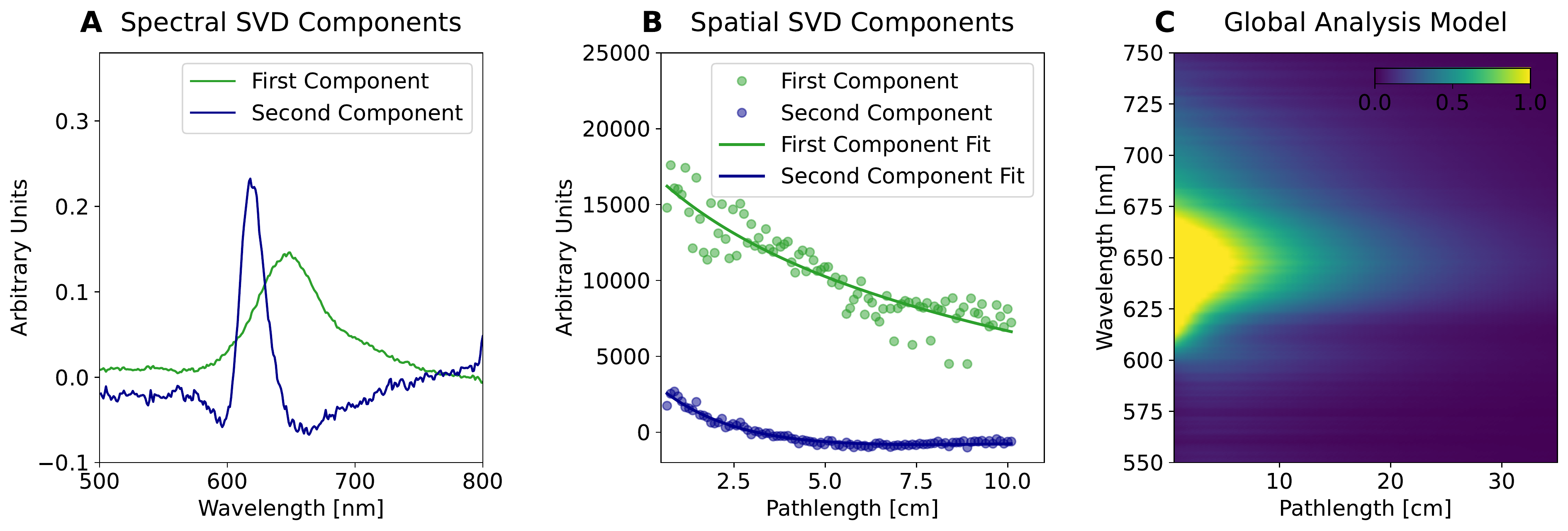}
  \caption{ \textbf{A}\&\textbf{B} - SVD decomposition of the spatially resolved photoluminescence for the perylene LSC, highlighting the major spectral and spatial components of the SVD analysis. Solid lines in B are the recovered fit using the presented analysis method. \textbf{C} - Analysis model extending the modelled edge photoluminescence for large LSCs.
   }
  \label{fig:3}
\end{figure}

We model spatial dependence of LSC photoluminescence by assuming the photoluminescence follows a sum of weighted exponentials, following Beer's law,

\begin{equation}\label{eq:global_analysis}
PL(\overline{l}, \lambda)=\sum_{i} A_{i}(\lambda) \ \mathrm{e}^{-\overline{l} \ \alpha_{i}},
\end{equation}

\noindent where $i$ corresponds to the size of the basis expansion used to describe the data, $A_{i}(\lambda)$ denotes the coefficient related to each wavelength and $\alpha_{i}$ is the absorption coefficient associated with each exponential decay. We conduct a simultaneous analysis of the photoluminescence spectra traces at all wavelengths by using singular value decomposition (SVD) where we reduce the spatially dependent photoluminescence data to its primary components.

SVD facilitates the interpretation of observed spatially resolved photoluminescence by reducing the dimensionality of the problem. Let matrix $\mathrm{D}$ describe the measured photoluminescence at the measured pathlengths where $\mathrm{D}$  is an $m \times n$ real matrix with $m>n$, where $m$ is the number of pathlengths sampled, $n$ is the number of wavelengths recorded, then $\mathrm{D}$ can be written in the  form
\begin{equation}
    \mathrm{D}=\mathrm{US} \mathrm{V}^{\mathrm{T}}.
\end{equation}
\noindent where $\mathrm{U}$ has dimensions of  $m \times m, \mathrm{S}$ has $m \times n,$ and $\mathrm{V}$ has $n \times n$. $\mathrm{U}$ and $\mathrm{V}$ are unitary, so that $\mathrm{U}^{\mathrm{T}} \mathrm{U}=\mathrm{I}$ and $ \mathrm{V}^{\mathrm{T}} \mathrm{V}=\mathrm{I}$ (where the two identity matrices may have different dimensions). $\mathrm{S}$ has entries only along the diagonal, known as singular values. The weighted left singular vectors (wLSV) are given by $\mathrm{US}$.

Decomposing the spatial data in this way has a useful interpretation; $\mathrm{U}$ is the matrix of left singular vectors giving the spatial dependence of the signal, $\mathrm{V}^{T}$ is the matrix giving the spectral dependence of the signal as plotted in \textbf{Figure \ref{fig:3}A} and \textbf{B}, respectively. The goal is to determine what subset of the data is required to adequately describe the full dataset. The best practice in choosing SVD components is to target a minimally descriptive model, using the smallest possible set of components to describe the data \cite{VanStokkum2004GlobalSpectra}.

 Adapting Beer's law (\textbf{Equation \ref{eq:global_analysis}}) into matrix form, and using the reduced weighted left singular vectors in place of the full data matrix, we can write
\begin{equation}\label{eq:matrix_global_analysis}
    (\mathrm{U S})_{n}=\mathrm{E}(\vec{\alpha}) \mathrm{x},
\end{equation}
where (US)$_{n}$ represents the matrix of chosen weighted left singular vectors. Here $\mathrm{E}$ is the design matrix, which is an exponential function of the absorption coefficients vector, $\vec{\alpha}$. Array $\mathrm{x}$ corresponds to the coefficient, $A_{i}$, for each weighted left singular vector. The problem becomes for what vector of absorption coefficients, $\vec{\alpha}$, is \textbf{Equation \ref{eq:matrix_global_analysis}} best satisfied, which can be solved efficiently using any good numerical solver, by solving the associated least squares problem,

\begin{equation}\label{eq:minimisation_problem}
\min _{ \vec{\alpha}}\left\|(U S)_{n}-E(\vec{\alpha}) x\right\|_{2}^{2},
\end{equation}

\noindent where the subscript refers to the Euclidean norm. The residue is then the norm of the square of all the differences. If the number of exponentials is not sufficient to describe the measured data, this suggests a number of absorption coefficients higher than the number of components detected by SVD.  The number of components need not be equal to the number of spectrally distinct components present \cite{Ruckebusch2012ComprehensiveReview}.

From the SVD analysis, we find two components are sufficient to describe the data in the case of the perylene red LSC. \textbf{Figure \ref{fig:2}C}  plots the extrapolated edge photoluminescence returned by the model to larger LSCs than measured, here up to a $25$ cm pathlength, corresponding to a $\sim 40 \text{ by} 40 \text{ cm}$ LSC. As we have now obtained the edge photoluminescence spectra for LSC of arbitrary size, we may now predict $\eta_{\text{int}}$ for arbitrary large LSCs.

\section{Results and Discussion}

\begin{figure}
\centering
  \includegraphics[width=0.95\linewidth]{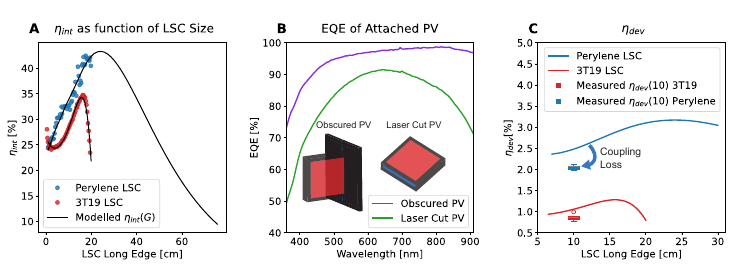}
  \caption{ \textbf{A} - $\eta_{\text{int}}$ from the measurements are by dots and the solid line is determined from the analysis model. \textbf{B} - External quantum efficiency (EQE) of two different solar cells. The green line is the solar cell which has been laser cut to match the edge of the LSC. The purple line is the solar cell fully intact but largely obscured other than an area to match the edge surface area of the LSC. \textbf{C} -  Solid lines give $\eta_{\text{dev}}$ determined from the analysis model, and data points are 5 similar LSCs coupled to the solar cells and $\eta_{\text{dev}}$ determined from I-V curves. %\hl{caution - need to redo perylene LSC EQE measurements} #figure needs changed to perylene
   }
  \label{fig:4}
\end{figure}

The measured and extrapolated $\eta_{\text{int}}$ as a function of LSC size  for the perylene red LSC and a standard 3T19 LSC (see \textbf{Methods \ref{methods:lsc_manufacture}} for details) are given \textbf{Figure \ref{fig:4}A}. Here, the recorded photoluminescence as a function of pathlength (\textbf{Equation \ref{eq:pathlength}}) has been corrected for the angle subtended (see \textbf{SI Section 1.8}), supposing that the illumination point is the centre of the imagined LSC. The measured $\eta_{\text{int}}$ (blue dots) and the modelled $\eta_{\text{int}}$ (black line), are plotted, with good agreement between the two. The black line also extends $\eta_{\text{int}}$ to LSC sizes far beyond what is practical to place into an integrating sphere. The model reproduces the characteristic inflection point observed in LSCs which exhibit self-absorption \cite{Currie2008High-efficiencyPhotovoltaics}. Notably, this is predicted by the model for the perylene red LSC even before the inflection is reached. This method may be readily extended to include the effect of back reflectors or mirrors.

Methods for experimentally determining $\eta_{\text{dev}}$ are well established in solar cell and LSC literature \cite{Yang2022ConsensusPerformance, Debije2021LaboratoryConcentrators,Yang2019HowConcentrators,Aste2015PerformanceModule,Waldron2017,Slooff2008AEfficiency}. However, significant variation of reported $\eta_{\text{dev}}$ exists in the literature for similar LSCs, as $\eta_{\text{dev}}$ is highly dependent on the nature of the attached solar cell \cite{Roncali2020LuminescentVadis,Yang2022ConsensusPerformance}.  Extreme care must be taken in $\eta_{\text{dev}}$ measurements to ensure that there is no direct illumination of the solar cells by the light source and to minimize reflection of light initially transmitted through the LSC. Coupling the solar cell to the LSC as well as identifying the active area also induces significant scope for systematic error which is difficult to determine. Various methods are used for attaching solar cells to an LSC, including refractive index matching optical tape, or index matching solutions and epoxy \cite{Currie2008High-efficiencyPhotovoltaics,Waldron2017}.

The external quantum efficiency of the system can be easily determined for arbitrary sized LSCs by adapting the model if the external quantum efficiency (EQE) of the solar cell is known (see \textbf{SI Section 1.10}). However, with a little effort and some assumptions, it is also possible to approximate $\eta_{\text{dev}}$ (\textbf{Equation \ref{eq:etadev}}) as a function of LSC size from the spatially resolved photoluminescence. The incident optical power on the LSC surface, assuming standard terrestrial illumination, is the integral of the terrestrial solar spectrum, $\text{AM1.5}(\lambda)$, in Watts per metre squared, over the active area of the LSC. The output power can be then calculated from the short circuit current which is integral of the EQE of the side-mounted solar cell, the edge photoluminescence corrected by some photon conservation factor, times $V_{\text{OC}}$ and $FF$,

\begin{equation} \label{eq:eqe2}
    \eta_{\text{dev}}(G) = \frac{ V_{\text{OC}} \ FF \overbrace{\int_{0}^{\infty} \lambda/hc \ \text{AM1.5}(\lambda)   \ \text{Abs}(\lambda) \ \text{d}\lambda \ \int_{0}^{\infty}  \ q  \ \eta_{\text{int}}(G) \  \text{PL}(G,\lambda) \ \text{EQE}(\lambda)   \ \text{d}\lambda}^{\propto \ I_{SC}}}{\int_{0}^{\infty}  \ \text{AM1.5}(\lambda) \ \text{d}\lambda},
\end{equation}

\noindent where $\text{PL}(G,\lambda)$ is the normalised photoluminescence $\left(\int_{0}^{\infty} \text{PL}(G,\lambda) \text{d}\lambda = 1 \right)$ in photons per second per nanometer. The absorption, $\text{Abs}(\lambda)$, is given by $\text{Abs}(\lambda) = 1-10^{-A_{\text{abs}}(\lambda)}$ where $A_{\text{abs}}$ is the absorbance.

The values to use for the open circuit voltage $V_{\text{OC}}$ and the $FF$ in \textbf{Equation \ref{eq:eqe2}} may be determined from either direct measurement or using a suitable diode model. Utilising an appropriate diode model, it is possible to relate the $V_{\text{OC}}$ and $FF$ of the solar cell measured directly under AM1.5, to the emitted photon flux and its spectra. However, in our case, we determined that $FF = 0.48 $ and $V_{\text{OC}} = 0.62 \ V$ by directly measuring from the PV attached to the emitting edge of the LSC (see \textbf{Methods \ref{methods:eqe}} for details). To use \textbf{Equation \ref{eq:eqe2}}, we make the explicit assumption here that the EQE nor $\eta_{int}$ are strongly dependent on the photon flux, although this may be relaxed by measurement of either as a function of power. However, more difficult to determine is the extent that the $FF$ and $V_{\text{OC}}$ change, which limits over what range \textbf{Equation \ref{eq:eqe2}} is valid. \textbf{Equation \ref{eq:eqe2}} supposes that all the light leaving the LSC will make it to the PV, whereas in fact some may be reflected at the PV interface or coupling optics. However, as long as changes in the photon flux are not much greater than $\pm 30$ \% this should make a negligible change to $FF$ or $V_{\text{OC}}$ chosen  (see \textbf{SI Section 1.11} for details). Particularly poor couplings of PV to LSCs should not use \textbf{Equation \ref{eq:eqe2}}.

\textbf{Equation \ref{eq:eqe2}} is plotted for the LSC in \textbf{Figure \ref{fig:4}C}, which represents a perfect case, neglecting coupling losses and concentration effects on the EQE of the PV cell.  We measured $\eta_{\text{dev}}$ for both LSCs using the taped solar cell. Comparing measurements made from IV measurements to the theoretical $\eta_{\text{dev}}$ reveals our coupling losses and voltage losses are as high as 15\% of $\eta_{\text{dev}}$.  Values of 20\% have been previously anticipated \cite{Yang2022ConsensusPerformance}. This highlights further difficulties in relating $\eta_{\text{dev}}$ directly to the optical performance of the LSC.

We highlight here the importance of providing the EQE as a function of wavelengths of the attached solar cell. \textbf{Figure \ref{fig:4}B} highlights the difference between where the solar cell has been laser cut to match the side edge of the LSC and where it has been taped to match the active area of the LSC. The laser treatment results in a decreased EQE (\textbf{Figure \ref{fig:4}}) compared to the PV where the active area has been taped to match the size of the LSC. If the EQE of the solar cell is not given, $\eta_{dev}$ says little to the effectiveness of the LSC.

\section{Conclusion}

Notwithstanding the method presented here, for full devices, where the LSC is coupled to PV, we believe it is vital that researchers carry out the standard reporting of $\eta_{\text{dev}}$, as only this figure will allow the community to track the meaningful impact of LSCs and allow for comparisons to the wider PV literature \cite{Yang2022ConsensusPerformance}. Further, without providing the EQE of the solar cell as a function of wavelength, it is impossible to deconvolve $\eta_{\text{dev}}$ and $\eta_{\text{int}}$. Although $\eta_{\text{dev}}$ remains the figure of merit, we caution the difficulties in using $\eta_{\text{dev}}$ as a design tool when considering the optical properties of the LSC as coupling the PV to the LSC may obfuscate $\eta_{\text{int}}$.

  Performance of the optical properties of the LSC are of paramount research interest in LSC research. As such $\eta_{\text{int}}(G)$ is valuable guide where to spend efforts to improve performance. We consider that the major advantage of the proposed method is that it provides an accurate means of determining both the $\eta_{\text{int}}$ and maximum potential $\eta_{\text{dev}}$ as a function for arbitrary LSC size and shape, within one system. The method allows the experimentalist to outline reasonable $\eta_{\text{int}}$ and $\eta_{\text{dev}}$ for large scale window sized LSCs, which cannot be realistically produced with typical laboratory facilities.  Using spatially resolved photoluminescence measurements is possible to visualise the losses, and easily determine efficiency benefits arising from different technologies, such as different back reflectors or optimise for potential improvements arising from different solar cell technologies, if their EQE is known. Further, it is trivial to control photon flux, which is of importance for future LSC technologies\cite{Erickson2019PhotoluminescenceDownconversion}. The global analysis method reported here has the advantage compared to previously reported spatially resolved methods that we need not approximate the self-absorption ratio or assume a single peak wavelength of emission \cite{Currie2008High-efficiencyPhotovoltaics, Batchelder1979,Batchelder1981LuminescentEfficiencies}. The authors hope that spatially resolved photoluminescence measurements may lead to the visualisation of more complex loss channels and provide future insights to improve LSC efficiency.

  \section{Methods}

\subsection{LSC Manufacture} \label{methods:lsc_manufacture}

We utilise two LSCs throughout this paper. In an effort to introduce an easily reproducible standard, we utilised a commercially available acrylic known as 3T19, often referred to as Lava Orange, which is manufactured by Lucite International and is often sold under the Perspex brand. The commercial LSC is widely available and relatively affordable, on the order of  2 USD per 10 cm$^{2}$. We purchased 3 LSCs from three resellers, which were laser cut and polished to form a 10 x 10 x 0.3 cm LSC. 3T19 was consistent across 3 different suppliers (see \textbf{SI Section 1.5.3}). The advantages of the 3T19 LSC are its reproducibility, robustness, longevity, ease of cleaning and ubiquity. However, although the dye concentration remains the same across the suppliers sampled, no published information is available on this value or the structure of the emitting dye. Therefore, to have control over the luminophore, we also manufactured an LSC using perylene red (CAS 123174-58-3, Tokyo Chemical Industries).

A stock solution of monomer was prepared by mixing 80\% lauryl methacrylate (96\%, 500 ppm MEHQ inhibitor, CAS 142-90-5, Merck) and 20\% ethylene glycol dimethacrylate (98\%, 90–110 ppm MEHQ inhibitor, CAS 97-90-5) with ($0.10 \pm 0.025$)\% UV initiator 2,2-dimethoxy-2-phenylacetophenone, CAS 24650-42-8, Merck) by weight under ambient atmosphere and degassing in a vacuum chamber. The mixtures were then placed in between two glass sheets with a PTFE spacer resulting in dimensions of $10$ cm by $10$ cm by $2$ mm. The mixture was then injected between the glass sheets, and exposed to $385$ nm LEDs (Wicked Engineering, CUREbox) for 5 minutes, before being left overnight in the dark. %The LSCs were then removed and used without any back reflector unless otherwise mentioned.

\subsection{Integrating Sphere Details} \label{methods:integratingsphere}

An optical fibre (Andor SR-OPT-8019) leads from the sphere to a grating spectrograph (Andor Kymera-328i) and detectors (Andor iDus 420 and iDus InGaAs 1.7). Immediately in front of the optical fibre port is a baffle, also coated with barium sulfate, preventing direct illumination of the optical fibre, and one-bounce illumination of the optic fibre. This arrangement sets geometric conditions on the size and placement of the baffle with respect to the size of the LSC (\textbf{SI Section 1.2}). The laser (Thorlabs L405G1, profile and stability details in \textbf{SI Section 1.3}) is coupled directly to the sphere and mounted on a temperature-controlled stage (Thorlabs LDM56). Coupling optics were supplied by Thorlabs and modified in-house to fit the ports of the integrating sphere.

\subsection{Solar Simulator and Power Measurement Details} \label{methods:eqe}
A solar simulator (Unisim, TS-SpaceSystems) was used which replicates AM1.5G. Silicon solar cells from  SunPower (California, United States) rated at 22\% efficiency were coupled to the LSC using refractive index matching tape (3M, USA). For demonstration purposes, we also used laser solar cells from Solar Made (Colorado Springs, USA), highlighting differences that the properties of the attached solar cells can make on reported $\eta_{\text{dev}}$. Diode characteristics of the PV cells are obtained by connecting them with gold Kelvin clips to a LabView-controlled Keithley 2400 Digital SourceMeter. The load was then varied to generate an I–V curve.

$\eta_{\text{dev}}$ was then determined from the ratio of the electric power from the side attached PV cell ($P_{LSC}$) to the incident power on the area of the LSC exposed to light ($P_{in}$), typically AM1.5,

\begin{equation} \label{eq:etadev}
\eta_{\text{dev}} = \frac{P_{\text{LSC}}}{P_{\text{In}}} = \frac{I_{\text{SC}}V_{\text{OC}}}{P_{\text{In}} }FF \text{ where }  FF = \frac{I_{\text{MP}}V_{\text{MP}}}{I_{\text{SC}}V_{\text{OC}}},
\end{equation}

\noindent and $I_{\text{SC}}$ is the short circuit current of the attached PV, $V_{\text{OC}}$ is the open circuit voltage, $FF$ is the fill factor and $V_{\text{MP}}$ $I_{\text{MP}}$ are the max power points.

\begin{acknowledgement}

 T.K.B. gives thanks to the Centre for Doctoral Training in New and Sustainable Photovoltaics for ﬁnancial support. We acknowledge financial support from the EPSRC and the Winton Program for the Physics of Sustainability. This project has received funding from the European Research Council (ERC) under the European Union’s Horizon 2020 research and innovation programme (Grant agreement no. 758826).

\end{acknowledgement}

\subsection*{Author Contributions}
TKB and AR conceived of the presented idea. TB developed the experimental arrangement and performed the computations presented. JX advised on the $\eta_{IQE}$ uncertainty analysis. BD assisted in the integrating sphere calibration and spectrometer alignment.  NCG and AR supervised the work.

The authors declare no competing financial interest.

\begin{suppinfo}

The supplementary information contains further information on integrating sphere design rules and calibration. It also includes detailed error propagation and mathematical derivations referred to in the text. All code for the simultaneous analysis is available on Github \hl{URL TO BE ADDED}. Measured data reported in the main text is available on the Cambridge Data Repository \hl{URL TO BE ADDED}.

\end{suppinfo}

\bibliography{references}

\end{document}